\documentclass[aps, prb, twocolumn, superscriptaddress, floatfix]{revtex4-2}

\usepackage[T1]{fontenc}
\usepackage[utf8]{inputenc}
\usepackage{amsmath,amssymb}
\usepackage{graphicx}
\usepackage{float}
\usepackage{booktabs}
\usepackage{longtable}
\usepackage{array}
\usepackage{calc}
\usepackage{tabularx}
\usepackage{url}
\graphicspath{{figures/}{../../figures/}}
\usepackage[hidelinks]{hyperref}

\makeatletter
\def\@bibdataout@rev{%
 \immediate\write\@bibdataout{@CONTROL{REVTEX42Control\eprint@enable@sw{}{,eprint="1"},note="1"}}%
 \if@filesw
  \immediate\write\@auxout{\string\citation{REVTEX42Control}}%
 \fi
 \@bibdataout@aps
}%
\makeatother

\begin{document}

\title{Accurate Molecular Property Regression Using Simple Fusion of Complementary Classical and Graph Predictors}
\author{\href{https://orcid.org/0000-0002-4910-2396}{Fei Yu}}
\email{mryufei@hku.hk}
\author{\href{https://orcid.org/0009-0005-1086-3605}{Jie Liu}}
\email{liujie@hkqai.hk}
\author{\href{https://orcid.org/0000-0001-5015-0902}{GuanHua Chen}}
\email{ghc@everest.hku.hk}
\author{\href{https://orcid.org/0000-0003-0862-5032}{Ziyang Hu}}
\email{hzy@yangtze.hku.hk}
\affiliation{Department of Chemistry, The University of Hong Kong, Pokfulam, Hong Kong SAR, China}
\affiliation{Hong Kong Quantum AI Lab Limited, Pak Shek Kok, Hong Kong SAR, China}

\begin{abstract}
Accurate prediction of continuous molecular properties supports chemical discovery and design. Many recent approaches combine multiple molecular representations within increasingly complex neural architectures, aiming to achieve higher accuracy. However, classical machine-learning models remain competitive when trained on chemically meaningful molecular representations. We therefore constructed a chemistry-guided classical branch from Morgan fingerprints, RDKit molecular descriptors, and conformer-derived geometric features selected to provide complementary views of molecular structure. We combined four tree learners trained on these representations with a CoMPT graph predictor and integrated their out-of-fold predictions using Ridge regression. Across ESOL, FreeSolv, and Lipophilicity datasets, the fused model achieved lower mean root-mean-square error (RMSE) than Chemprop and competitive graph and multimodal methods, including CoMPT, MoleSG, KFLM2, and MCMPP. After retraining on the separately curated AqSolDB dataset, the same approach, without redesign or retuning, reduced mean absolute error (MAE) relative to the classical ensemble and achieved lower mean MAE than Chemprop. These results show that accurate and transferable molecular property regression does not necessarily require a complex joint multimodal architecture: classical learning remains competitive when chemical insight guides the choice of molecular representations, and a graph model can add complementary information through simple prediction-level fusion.
\end{abstract}

\maketitle

\section{Introduction}\label{introduction}

Molecular property prediction supports chemical discovery and design.
Experimental measurements and high-fidelity calculations can be too slow
or costly to evaluate every candidate in large chemical spaces. Alternatively, molecular
property regression provides a computational route from molecular structure
to continuous properties and can help prioritize compounds for further study
\citep{wu2018moleculenet, yang2019chemprop, huang2021tdc}.

Molecular properties reflect chemical information expressed at several
structural scales. Relevant information may reside in local atom environments,
whole-molecule composition and physicochemical attributes, three-dimensional
conformation, and atom-bond connectivity. Molecular representations expose
different parts of this information to a learning algorithm. Fixed
representations calculate features from molecular structure using predefined
rules, whereas learned representations adapt features to a task or pretraining
objective \citep{xia2023multitask}. No single representation is guaranteed to
make all relevant information accessible to a predictor. A useful
multimodal design therefore depends on both the chemical information
exposed by each representation and the way those representations are
integrated.

Model families differ in both the molecular information they receive and
the relationships they can learn. Classical algorithms fit fingerprints
and molecular descriptors; graph neural networks learn from atoms and
bonds; molecular sequence models learn from strings such as SMILES, and
language-model approaches use pretrained chemical or general-purpose
text representations
\citep{chen2016xgboost, prokhorenkova2018catboost, ke2017lightgbm, geurts2006extratrees, chen2021compt, ross2022molformer, jablonka2024predictivechemistry}.
Guo and Ding compared these families across 26 drug-discovery endpoints
and found that model scale alone did not determine predictive
performance. Classical fingerprint-based and descriptor-based learners were
the strongest family overall, while graph and pretrained sequence models
gained or lost ground depending on the endpoint and data split
\citep{guo2026larger}. Their benchmark concerns model-family performance
rather than fusion, but it shows why classical learners should remain
substantive predictors and strong baselines.

Several multimodal methods combine information across representations.
MoleSG jointly processes learned sequence and graph representations,
KFLM2 combines domain-tuned language and graph representations through
several fusion strategies, and MCMPP uses cross-attention across text,
sequence, graph, and geometry
\citep{shen2024molesg, xie2025kflm2, sun2025mcmpp}. Parker and
co-workers combined graph-derived features, general molecular
descriptors, and heterogeneous learners \citep{parker2025heterogeneous}.
These studies establish that representation diversity can be useful, but
they also leave open how much additional architecture is needed to
integrate strong predictors.

Here, we ask whether chemistry-guided representation selection and
simple fusion of strong classical and graph predictors can deliver
accurate molecular property regression without constructing another
joint multimodal architecture. We selected Morgan fingerprints, RDKit
molecular descriptors, and conformer-derived geometric features to
expose complementary information about local substructures,
whole-molecule physicochemical attributes, and conformer geometry.
CoMPT provides a separate atom-bond graph view \citep{chen2021compt}.
Ridge regression combines the five prediction channels after training
on grouped out-of-fold predictions
\citep{wolpert1992stacking, hoerl1970ridge, varma2006cvbias}. These
steps implement the choice to keep representation-specific predictor
training separate from integration.

We evaluated the fused predictor on ESOL, FreeSolv, and Lipophilicity,
three established benchmarks for solution-phase molecular properties
\citep{delaney2004esol, mobley2014freesolv, wu2018moleculenet}. We
compared it with Chemprop on the same molecules and data splits and with
competitive graph and multimodal methods that reported results on the
same three benchmarks, including CoMPT, MoleSG, KFLM2, and MCMPP. We
selected them because CoMPT and MoleSG reported state-of-the-art
performance across multiple tasks, while KFLM2 and MCMPP reported
competitive results against strong baselines
\citep{chen2021compt, shen2024molesg, xie2025kflm2, sun2025mcmpp}. We
then retrained the unchanged configuration on AqSolDB, a larger,
separately curated aqueous-solubility dataset \citep{sorkun2019aqsoldb}.
This design tests both benchmark accuracy and whether the same simple
integration strategy remains useful on a new molecular collection.

\section{Methods}\label{methods}

\subsection{Study design and
datasets}\label{study-design-and-datasets}

We evaluated molecular property regression as a mapping from a molecular
structure \(m\) to a continuous target \(y\). The public benchmark suite
comprised ESOL with 1,128 aqueous-solubility measurements, FreeSolv with
642 experimental hydration free energies, and Lipophilicity with 4,200
log \(D\) measurements at pH 7.4
\citep{delaney2004esol, mobley2014freesolv, bento2014chembl, wu2018moleculenet}.
These benchmarks cover different solution-phase properties and dataset
sizes. Ten predefined random splits assigned molecules to the training,
validation, and test sets for each dataset. The exact assignments and
record counts are provided in the Supporting Information.

AqSolDB was used as a separate transfer dataset because it is a larger,
separately curated aqueous-solubility collection with a predefined test
split \citep{sorkun2019aqsoldb}. The full dataset contains 9,982 log \(S\)
records, including 1,997 test records. We excluded test molecules that
shared an exact structure or parent molecular connectivity with any
molecule used during earlier method development, as specified in Section
S7, or with any molecule in the AqSolDB training and validation data. We
also removed duplicate parent structures within the test set and
molecules for which any evaluated method could not generate its required
representation. After these filters, the overlap-controlled AqSolDB test
set contained 1,357 molecules and 474 distinct nonempty molecular
scaffolds. All five training and validation splits and all evaluated
methods used these same 1,357 test molecules. We fixed the model
configuration, test-set composition, data splits, evaluation metrics, and
summary procedure before examining the solubility values.

\begin{figure*}[t]
\centering
\includegraphics[width=\textwidth, trim=0 78 0 0, clip]{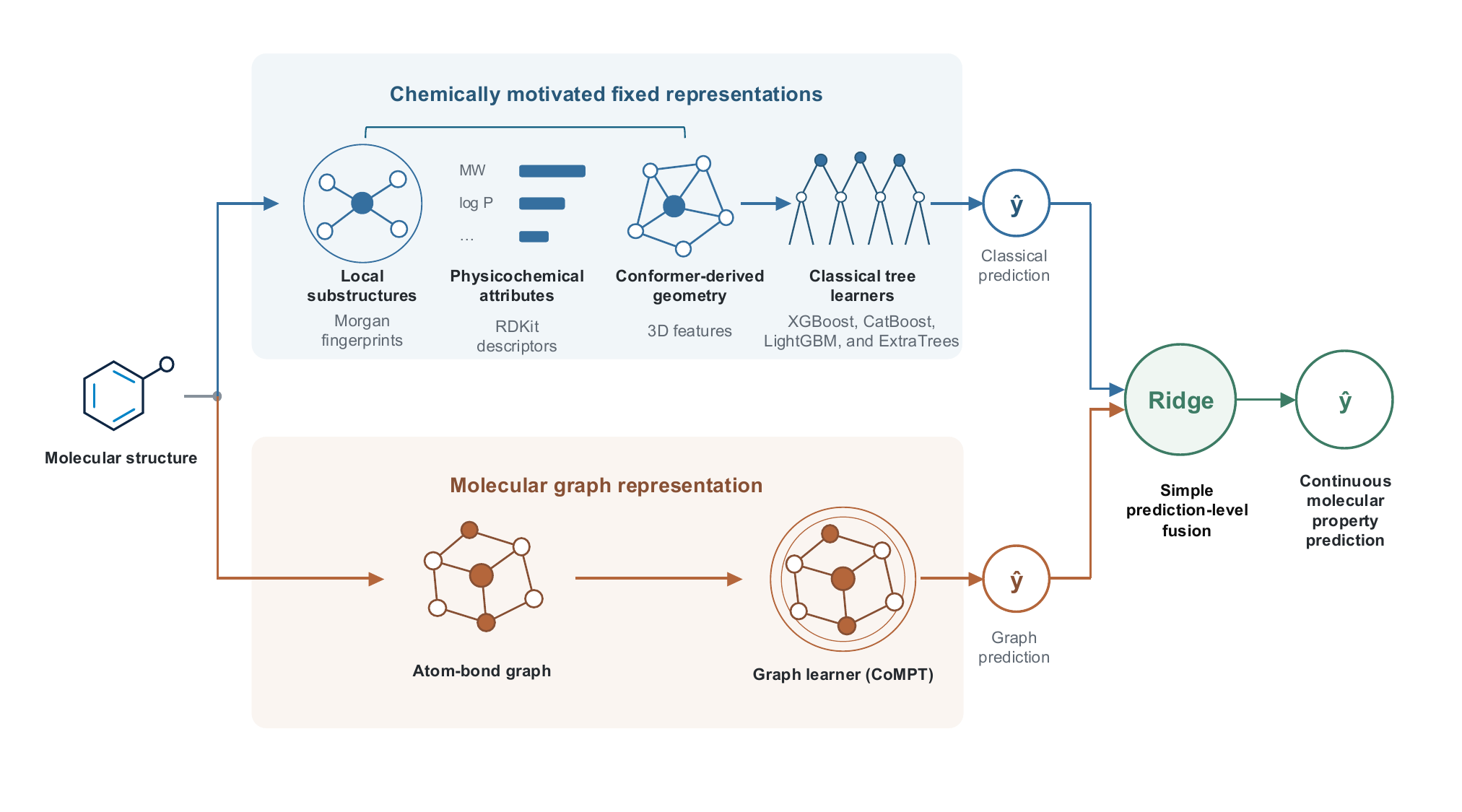}
\caption{Conceptual overview of complementary prediction-level fusion. A
molecular structure is encoded through chemically motivated fixed
representations that describe local substructures, whole-molecule
physicochemical attributes, and conformer-derived geometry, and separately as
an atom-bond graph. Classical tree learners and CoMPT produce distinct property
predictions, which Ridge regression combines at the prediction level. The Ridge
model is fitted to out-of-fold predictions.
}
\label{fig:workflow}
\end{figure*}

\subsection{Method overview}\label{method-overview}

The method has two prediction branches (Fig.~\ref{fig:workflow}). XGBoost, CatBoost,
LightGBM, and ExtraTrees learn from the same fixed inputs: Morgan
fingerprints, RDKit descriptors, and conformer-derived geometric
features. CoMPT learns separately from the molecular graph. Ridge
regression first combines the four tree predictions to form \(T\). A
second Ridge model combines the four tree predictions with the CoMPT
prediction \(G\) to form \(T\oplus G\). The symbol \(\oplus\) denotes
fusion at the prediction level.

The predictors remain separate until their outputs reach Ridge. This
design lets the tree models and graph model process their respective
inputs without forcing the underlying features into one shared
representation. To train Ridge without using in-sample predictions, each
training molecule received one prediction from every relevant base model,
and each prediction came from a model fitted on the other two folds. We
concatenated the predictions from the three held-out folds in the original
molecule order and fitted Ridge to the resulting matrix. The same
procedure defines both \(T\) and \(T\oplus G\).

\subsection{Molecular
representations}\label{molecular-representations}

The fixed inputs were selected to expose complementary chemical
information at several structural scales. A 2048-bit radius-2 Morgan
fingerprint without chirality records recurring
atom-centered neighborhoods extending up to two bonds from each atom
\citep{rogers2010ecfp}. The 217 RDKit descriptors provide whole-molecule
summaries, including molecular weight, polarity and lipophilicity
estimates, atom and ring counts, topological indices, and fragment
counts \citep{rdkit2026}.

To add spatial information, we calculated conformer-derived features
that describe molecular shape and the relative positions of chemical
groups. Because tree learners require a fixed number of input variables,
we summarized up to 16 deterministic ETKDGv3 conformers per canonical
molecule in a 317-dimensional vector
\citep{riniker2015etkdg, wang2020etkdgv3}. The descriptors cover global
three-dimensional shape, distances among donor, acceptor, aromatic,
hydrophobic, positive-ionizable, and negative-ionizable features,
donor-acceptor contacts, conformer diversity, the number of retained
conformers, and whether conformer generation required a retry. This
study-specific feature block is called Z3D in the implementation and
Supporting Information. The choice of 16 conformers and the 317-variable
schema were fixed design choices rather than a claim of chemical
optimality.

Some geometric descriptors are undefined when a molecule lacks the
corresponding pharmacophore type or when conformer generation does not
yield the required geometry. We therefore paired the 317 geometric
values with 317 binary indicators that distinguish observed values from
substituted values. The geometric features were calculated solely from
molecular structure and did not depend on the measured property. Within
each out-of-fold fit, missing or nonfinite values were imputed using
statistics estimated only from the fitting molecules, and the fitted
imputer was applied to the held-out molecules. The Supporting
Information gives the rule used when an entire feature lacked a finite
fitting value. For test prediction, the imputer was fitted on the
complete training set and applied to the test molecules.

In contrast to these predefined features, CoMPT learns atom-level and
bond-level representations from the molecular graph and converts them
into a property-specific prediction \citep{chen2021compt}. Its encoder
uses communicative message passing together with pairwise molecular
information. We included CoMPT because graph message passing may
preserve atom-bond relationships that are not represented explicitly by
fingerprints, molecular descriptors, or summary statistics of
conformers.

\subsection{Base prediction models}\label{base-prediction-models}

The concatenated fixed representation was supplied to XGBoost, CatBoost,
LightGBM, and ExtraTrees
\citep{chen2016xgboost, prokhorenkova2018catboost, ke2017lightgbm, geurts2006extratrees}.
The three boosting methods use different tree-construction and
regularization strategies, while ExtraTrees introduces stronger
randomization. We used these models to obtain different nonlinear
predictions from the same chemical inputs. Model configurations were
held constant in the reported comparisons; complete hyperparameters are
listed in the Supporting Information.

The local CoMPT implementation was verified against the reference
implementation as described in the Supporting Information. The
architecture contained three communicative encoder layers, a
bidirectional gated recurrent unit readout, and a regression head.
Within each out-of-fold fitting set, a validation subset containing
different canonical molecules was used to select the training epoch with
the lowest validation RMSE. The final graph model was then fitted on the
complete training set for test prediction.

The graph branch required a separate size constraint. CoMPT represents
each molecule in tensors padded to a validated maximum number of atoms.
ESOL and FreeSolv used the original 100-atom configuration.
Lipophilicity required a 115-atom configuration to include the largest
retained molecule, and this configuration was carried unchanged into the
AqSolDB evaluation. Molecules exceeding the applicable limit were
excluded rather than truncated. The Supporting Information reports the
architecture, optimization settings, and exclusions associated with this
rule.

\subsection{Out-of-fold fusion training and test
inference}\label{out-of-fold-fusion-training-and-test-inference}

Training followed four steps. First, we divided the training molecules
into three folds. Repeated records of the same standardized structure,
if present, remained in one fold. Second, each base model was fitted on
two folds and used to predict the third. Third, we placed the held-out
predictions for all training molecules into one matrix and fitted the
Ridge combiner. Finally, we refitted every base model on all training
molecules and applied the already fitted combiner to their test
predictions.

The following equations formalize this procedure.

Let \(\mathcal D_{\mathrm{tr}}=\{(m_i,y_i)\}_{i=1}^{n}\) denote the
training data. We partitioned the training molecules into \(K=3\) folds.
Fold assignment was grouped by canonical SMILES, so repeated records of
the same standardized structure could not occur in both the fitting and
held-out subsets. For fold \(k\), let \(\mathcal I_k\) and
\(\mathcal J_k\) denote the fitting and held-out index sets.

Let \(c\) index one of the four tree regressors or the CoMPT graph
model, and let \(A_c\) denote its learning procedure. For each fold
\(k\), we fitted \(\hat f_{c,k}\) using only the molecules indexed by
\(\mathcal I_k\) and predicted only the held-out molecules indexed by
\(\mathcal J_k\): \[
\hat f_{c,k}=A_c\!\left(\{(m_j,y_j):j\in\mathcal I_k\}\right), 
o_{ic}=\hat f_{c,k}(m_i), i\in\mathcal J_k.
\]

For a fixed-representation predictor, \(A_c\) includes the fold-specific
geometric-feature imputation described above. Collecting the held-out
predictions from the four tree regressors gives
\(O_T\in\mathbb R^{n\times4}\). Appending the CoMPT prediction gives
\(O_{T\oplus G}\in\mathbb R^{n\times5}\). For either model
\(r\in\{T,T\oplus G\}\), let \(O_r\) denote its out-of-fold prediction
matrix and \(p_r\) its number of columns.

For each column \(c\) of \(O_r\), let \(\mu_{r,c}\) and \(s_{r,c}\) be
the mean and standard deviation estimated from that column, with
\(s_{r,c}=1\) when its variance is zero. The standardized matrix has
entries \(\widetilde o_{r,ic}=(o_{r,ic}-\mu_{r,c})/s_{r,c}\). Let
\(\boldsymbol\beta_r\) denote the coefficients assigned to the
standardized base-model predictions and \(b_r\) the intercept. The Ridge
combiner for model \(r\) is
\begin{align*}
(\widehat{\boldsymbol\beta}_r,\widehat b_r) &=
\arg\min_{\boldsymbol\beta_r,b_r}
\left\|\mathbf y-\widetilde O_r\boldsymbol\beta_r-b_r\mathbf 1\right\|_2^2
+\lambda\|\boldsymbol\beta_r\|_2^2, \\
\lambda &=10.
\end{align*}
Here, \(\mathbf 1\) is the length-\(n\) vector of ones.

For test inference, each base predictor was refitted on
\(\mathcal D_{\mathrm{tr}}\) to give \(\hat f_c^{(\mathrm{all})}\). For
a test molecule \(m^*\), model \(r\) predicts \[
\widehat y_r(m^*)=
\widehat b_r+
\sum_{c=1}^{p_r}\widehat\beta_{r,c}
\frac{\hat f_c^{(\mathrm{all})}(m^*)-\mu_{r,c}}{s_{r,c}}.
\] The out-of-fold means, scales, and Ridge coefficients were applied
without refitting on the test set.

\subsection{Language-model representation
controls}\label{language-model-representation-controls}

We tested frozen hidden states from Qwen3-1.7B as auxiliary molecular
representations to ask whether they could provide information beyond the
fixed descriptors and graph model \citep{yang2025qwen3}. The input
combined canonical SMILES, a property-prediction task statement, and a
structured summary of conformer-derived features. We mean-pooled four
2,048-dimensional views spanning middle and final transformer layers and
evidence and global token regions. Within each fitting fold, principal
component analysis reduced each view to 32 dimensions before the
representations were used as direct regression features, residual
predictors, or additional out-of-fold predictions.

We used three controls across these exploratory experiments. (1) The
shuffled-state control reassigned hidden states among molecules, breaking
the correspondence between a molecule and its representation while
preserving the collection of hidden-state vectors. This tested whether
any gain required molecule-specific alignment. (2) The content-empty
control used hidden states generated from prompts without molecular
content. It retained the language-model extraction procedure while
removing the molecular information supplied in the prompt. This tested
whether the procedure could yield a gain without molecule-specific
chemical content. (3) The no-hidden control used the otherwise matched
downstream predictor without hidden-state inputs. It tested whether
adding the hidden-state representation improved the existing model. The
exact control paired with each integration experiment is specified in the
Supporting Information, which also gives the model revision, extraction
procedure, and complete control matrix.

\subsection{Reference methods, evaluation metrics, and statistical
summaries}\label{reference-methods-evaluation-metrics-and-statistical-summaries}

Chemprop v2.3.1 provided a learned molecular baseline evaluated on the
same molecules and data splits
\citep{yang2019chemprop, graff2026chempropv2}. It used a
bond-message-passing model with mean aggregation. For each fit, we
retained the training epoch with the lowest validation root-mean-square
error. Published CoMPT, MoleSG, KFLM2, and MCMPP results were compared
under the evaluation settings reported by their authors
\citep{chen2021compt, shen2024molesg, xie2025kflm2, sun2025mcmpp}.

Root-mean-square error (RMSE) was the evaluation metric for ESOL,
FreeSolv, and Lipophilicity. For Fig.~\ref{fig:rmse-deltas}, we calculated the reduction in
RMSE relative to each component as
\(\operatorname{RMSE}(T)-\operatorname{RMSE}(T\oplus G)\) and
\(\operatorname{RMSE}(G)-\operatorname{RMSE}(T\oplus G)\). Positive
values indicate lower error after fusion. AqSolDB used mean absolute
error (MAE) as the primary metric and RMSE as the secondary metric. Its
paired differences were calculated as fused minus reference error, so
negative values favor the fused model.

For each dataset, we calculated means and sample standard deviations
across the split-specific results and counted the splits in which fusion
lowered the error. Results were summarized separately because the
properties have different units. For AqSolDB, we calculated a 95\%
Student-\(t\) interval from the five split-specific MAE differences.
Because the five fits used the same test molecules, this interval
summarizes sensitivity to the training and validation partitions rather
than variation among independent test samples.

\section{Results and Discussion}\label{results-and-discussion}

\subsection{The graph predictor complements the classical
ensemble}\label{fusion-improves-the-fixed-representation-ensemble}

\begin{figure*}[t]
\centering
\includegraphics[width=\textwidth]{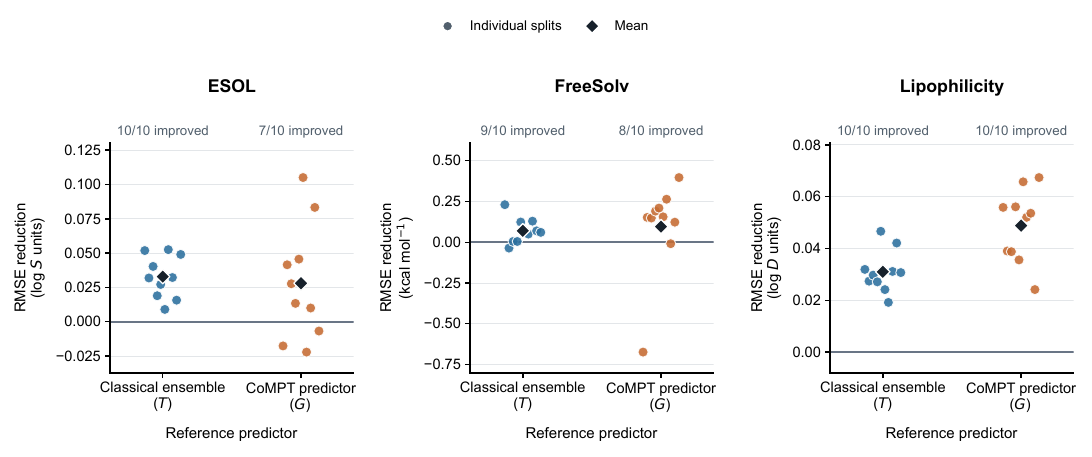}
\caption{RMSE reduction achieved by fusion across the three public
benchmarks. Each panel shows the reduction relative to the classical
ensemble \(T\) and the CoMPT predictor \(G\) across ten predefined data
splits. Positive values indicate lower error after fusion. Points
represent individual splits, diamonds represent the mean, and labels
give the number of splits in which fusion lowered RMSE. Source data are
provided in Table S5.
}
\label{fig:rmse-deltas}
\end{figure*}

Fusion lowered mean RMSE relative to both component predictors on every
public benchmark (Table~\ref{tab:public-benchmarks}). Mean \(T\oplus G\) RMSE was 0.550 for ESOL,
0.907 kcal mol\(^{-1}\) for FreeSolv, and 0.541 log \(D\) units for
Lipophilicity. Relative to \(T\), the corresponding mean reductions were
0.033, 0.070 kcal mol\(^{-1}\), and 0.031 log \(D\) units;
fusion lowered RMSE in 10, 9, and 10 of the respective ten splits.
Relative to \(G\), the mean reductions were 0.028, 0.096 kcal
mol\(^{-1}\), and 0.049 log \(D\) units, with lower RMSE in 7, 8,
and 10 splits, respectively (Fig.~\ref{fig:rmse-deltas}).

For the comparison with \(T\), the only increase occurred in FreeSolv
seed 29, where RMSE rose from 0.904 to 0.940. Removing the most favorable
split from each dataset did not reverse the mean improvement. FreeSolv
showed greater split-to-split variability than the other benchmarks
because one 65-molecule test split produced high errors for all evaluated
models. For \(T\oplus G\), the ten-split median RMSE was 0.710 and the
range was 0.566 to 2.524; all split-level values and split identifiers
are reported in Table S5.

Neither component predictor dominated the other across every split, yet
\(T\oplus G\) achieved lower mean RMSE than both on all three datasets.
This pattern is consistent with the CoMPT predictor \(G\) retaining
predictive information that the fixed-representation branch did not
fully capture. Calibration, variance reduction, and Ridge regularization
may also contribute, so the result supports the combined predictor
rather than a single mechanism of improvement.

\subsection{Comparison with reference
methods}\label{comparison-with-reference-methods}

On the same molecules and data splits, \(T\oplus G\) had lower mean RMSE
than Chemprop: 0.550 versus 0.667 for ESOL, 0.907 versus 1.404 kcal
mol\(^{-1}\) for FreeSolv, and 0.541 versus 0.575 log \(D\) units for
Lipophilicity \citep{yang2019chemprop, graff2026chempropv2}.

Published random-split results provided a broader comparison with
competitive graph and multimodal methods. Across five random-split runs,
KFLM2 reported mean RMSE values of 0.668, 1.079, and 0.821 for ESOL,
FreeSolv, and Lipophilicity, respectively. The corresponding published
CoMPT values were 0.562, 0.945, and 0.562; the
MoleSG values were 0.574, 0.924, and 0.580
\citep{xie2025kflm2, chen2021compt, shen2024molesg}. CoMPT includes
architectural and training hyperparameters that are typically tuned for
individual datasets. In this study, we did not tune CoMPT separately for
the three benchmarks but retained one fixed core configuration for
\(G\). This choice helps explain why the standalone \(G\) RMSE values in
Table~\ref{tab:public-benchmarks} were higher than the published CoMPT values. The corresponding
\(T\oplus G\) means were lower than KFLM2, CoMPT and MoleSG on all three tasks.

Table 2 of the MCMPP paper \citep{sun2025mcmpp} reports standard random-split RMSE values of 0.566,
1.032, and 0.611 for the same three tasks, which are also greater than the corresponding
\(T\oplus G\) mean values. Table~\ref{tab:public-benchmarks} collects the matched
Chemprop results and the published reference values.

\begin{table}[H]
\centering
\caption{Performance on the three public molecular-property benchmarks.}
\label{tab:public-benchmarks}
\footnotesize
\begin{tabularx}{\linewidth}{@{}>{\raggedright\arraybackslash}Xrrr@{}}
\toprule
Method & ESOL RMSE & FreeSolv RMSE & Lipophilicity RMSE \\
\midrule
\(T\) & \(0.583 \pm 0.071\) & \(0.977 \pm 0.639\) & \(0.572 \pm 0.022\) \\
\(G\) & \(0.578 \pm 0.080\) & \(1.003 \pm 0.343\) & \(0.590 \pm 0.027\) \\
\(T\oplus G\) & \(\mathbf{0.550 \pm 0.071}\) & \(\mathbf{0.907 \pm 0.583}\) & \(\mathbf{0.541 \pm 0.022}\) \\
Chemprop D-MPNN & \(0.667 \pm 0.105\) & \(1.404 \pm 0.173\) & \(0.575 \pm 0.026\) \\
\midrule
CoMPT & \(0.562 \pm 0.020\) & \(0.945 \pm 0.133\) & \(0.562 \pm 0.012\) \\
MoleSG & \(0.574 \pm 0.030\) & \(0.924 \pm 0.081\) & \(0.580 \pm 0.020\) \\
KFLM2 & \(0.668 \pm 0.049\) & \(1.079 \pm 0.069\) & \(0.821 \pm 0.012\) \\
\bottomrule
\end{tabularx}

\medskip
\raggedright\emph{Note.} Values from this study and Chemprop are mean \(\pm\) sample SD across ten predefined splits; the standard deviation describes variation among splits rather than among individual molecules. Chemprop used the same molecules and splits. CoMPT, MoleSG, and KFLM2 values are five-run means \(\pm\) sample SD reported in KFLM2 Supporting Information Table S6. ESOL errors are reported for \(\log_{10}[S/(\mathrm{mol\,L}^{-1})]\), FreeSolv errors in kcal mol\(^{-1}\), and Lipophilicity errors in log \(D\) units. Complete split-level values from this study are provided in Table S5.
\end{table}

\subsection{Transfer to AqSolDB}\label{transfer-to-aqsoldb}

AqSolDB was used to test whether the graph-fusion strategy remained
effective on a larger, separately curated aqueous-solubility dataset.
All models were retrained using AqSolDB data, but the molecular
representations, model classes, hyperparameters, and fusion procedure
were kept unchanged. We used five predefined training and validation
splits and evaluated every model on the overlap-controlled test set of
1,357 molecules described in Section \ref{study-design-and-datasets}.

Across the five AqSolDB splits, \(T\oplus G\) achieved the lowest mean
MAE and RMSE among the three study models. Relative to \(T\), mean MAE
decreased from 0.801 to 0.761, corresponding to an improvement of
5.0\%. The mean paired MAE difference was \(-0.040\), with a 95\%
Student-\(t\) interval of \([-0.059, -0.021]\), while mean RMSE decreased
from 1.081 to 1.039. Although \(G\) alone was less accurate than \(T\) on
AqSolDB, adding \(G\) improved both metrics, consistent with
complementary predictive information.

On the same test molecules, \(T\oplus G\) also showed better mean
performance than Chemprop D-MPNN. It achieved a lower MAE in four of the
five splits and a lower RMSE in all five. The mean MAE and RMSE of
Chemprop were 0.781 and 1.103, respectively, compared with 0.761 and
1.039 for \(T\oplus G\). The complete results are summarized in Table \ref{tab:aqsoldb-transfer}.

\begin{table}[H]
\centering
\caption{Performance on the overlap-controlled AqSolDB transfer test set.}
\label{tab:aqsoldb-transfer}
\footnotesize
\begin{tabularx}{\linewidth}{@{}>{\raggedright\arraybackslash}Xrr@{}}
\toprule
Method & MAE, mean \(\pm\) sample SD & RMSE, mean \(\pm\) sample SD \\
\midrule
\(T\) & \(0.801 \pm 0.015\) & \(1.081 \pm 0.017\) \\
\(G\) & \(0.833 \pm 0.053\) & \(1.123 \pm 0.052\) \\
\(T\oplus G\) & \(\mathbf{0.761 \pm 0.022}\) & \(\mathbf{1.039 \pm 0.019}\) \\
Chemprop D-MPNN & \(0.781 \pm 0.011\) & \(1.103 \pm 0.023\) \\
\bottomrule
\end{tabularx}

\medskip
\raggedright\emph{Note.} Values are the mean \(\pm\) sample standard
deviation across five predefined training and validation splits. All methods
were evaluated on the same 1,357 test molecules. MAE and RMSE are reported in
log \(S\) units. MAE was the primary evaluation metric and RMSE was secondary.
\end{table}

\subsection{Model specialization and language-model
controls}\label{model-specialization-and-language-model-controls}

The results support treating the fixed-representation learners and graph
model as specialists rather than competing alternatives. CoMPT did not
need to be the best standalone predictor to improve the combined model.
The experiments evaluate this complete configuration; because we did not
remove the feature blocks and tree learners one at a time, they do not
assign the observed gain to any single component.

The language-model experiments asked whether frozen Qwen3-1.7B hidden
states could add useful molecule-specific information. We evaluated the
compressed representations as direct regression features, predictors of
residuals from existing models, and additional out-of-fold predictions.

The experiments did not show a reproducible gain from correctly matched
hidden states. In the initial exploratory comparison, the shuffled
hidden-state control outperformed the correctly matched representation in
the residual-prediction setting. Across seven additional development
splits per task, the content-empty and no-hidden controls matched or
exceeded the correctly matched representation. We therefore excluded
hidden states from the final model under the tested representations and
integration procedures. It should be noted that, KFLM2 tests a different
setting in which language representations are domain-tuned and
incorporated within a jointly designed multimodal network
\citep{xie2025kflm2}.

\section{Conclusion}\label{conclusion}

Across the tasks studied, accurate multimodal molecular regression did
not require another joint neural fusion architecture. Chemistry-guided
selection of fixed representations supplied classical tree regressors
with complementary views of local substructure, whole-molecule
physicochemical attributes, and conformer geometry. The CoMPT supplied
additional atom-bond graph information as a specialized predictor. Their
statistical fusion reduced mean RMSE on ESOL, FreeSolv, and
Lipophilicity and achieved lower mean error than Chemprop on the same
molecules and splits. After retraining on AqSolDB without redesign or
retuning, the same configuration retained its advantage over the
fixed-representation ensemble and Chemprop on the overlap-controlled
test set. The results support a design strategy in which chemical
insight guides the choice of complementary representations. Moreover,
model sophistication and predictive strength need not increase together: strong
representation-specific predictors can remain separate and still deliver
accurate, transferable predictions when their outputs are integrated by
a simple statistical method.

\section{Associated Content}\label{associated-content}

\subsection{Supporting Information}\label{supporting-information}

Expanded methods, fixed hyperparameters, dataset and split identities,
the 30-case public-benchmark table, the five-seed AqSolDB table,
comparator sources, language-model representation controls, and software
and reproducibility information.

\subsection{Data Availability}\label{data-availability}

The graph-fusion benchmark and transfer evaluation used ESOL, FreeSolv,
Lipophilicity, and AqSolDB
\citep{delaney2004esol, mobley2014freesolv, wu2018moleculenet, sorkun2019aqsoldb}.
Earlier language-model control experiments described in the Supporting
Information also used Caco2\_Wang and LD50\_Zhu from Therapeutics Data
Commons \citep{huang2021tdc}. ESOL, FreeSolv, and Lipophilicity are
available from the cited public sources. The code repository provides
their upstream download locations and SHA-256 digests and reconstructs
the predefined benchmark splits from the upstream row order and recorded
random seeds. AqSolDB is available from its original public data deposit.
The repository does not redistribute molecular datasets, target values,
row-level predictions, or the overlap-controlled AqSolDB split
memberships. Exact reruns of the AqSolDB evaluation therefore require a
user-prepared dataset with the study partition columns. Public
configurations, exact software versions, source hashes, and aggregate-only
verification results are available at
\url{https://github.com/VCERS/molecular-property-fusion}. Third-party data
retain their original terms.

\subsection{Code Availability}\label{code-availability}

The code for feature generation, \(T\), \(G\), and \(T\oplus G\)
evaluation, the Chemprop baseline, AqSolDB transfer, summarization, and
aggregate verification is available in Molecular Property Fusion version
0.1.0 at
\url{https://github.com/VCERS/molecular-property-fusion} (commit
\texttt{4a2d86e}). The repository is released under the MIT License.
The repository records source commits and file hashes in a machine-readable
manifest. The CoMPT-derived graph implementation retains the upstream MIT
license and attribution, while Chemprop remains an external dependency at
version 2.3.1. Third-party components retain their original licenses.

\section*{Competing Interests}\label{competing-interests}

The authors declare no competing interests.

\section*{Acknowledgments}\label{acknowledgments}

This work was supported by Hong Kong Quantum AI Lab Limited, AIR@InnoHK of the Hong Kong Government, the GuangDong Basic and Applied Basic Research Foundation (Grant No. 2024A1515013283), and the National Natural Science Foundation of China (Grant No. 22473022).

\bibliography{references}

\end{document}